\documentclass[manuscript]{acmart}
\usepackage{tabularray}
\usepackage{makecell}
\usepackage{hyperref}
\usepackage{url}
\AtBeginDocument{%
  \providecommand\BibTeX{{%
    \normalfont B\kern-0.5em{\scshape i\kern-0.25em b}\kern-0.8em\TeX}}}

\begin{document}

\title[More Than Just Warnings:\\Exploring the Ways of Communicating Credibility Assessment on Social Media]{More Than Just Warnings:~Exploring the Ways of Communicating Credibility Assessment on Social Media}



\author{Huiyun Tang}
\affiliation{%
  \institution{University of Luxembourg,}
  \city{Esch-sur-Alzette}
  \country{Luxembourg}}
\email{huiyun.tang@uni.lu}

\author{Björn Rohles}
\affiliation{%
  \institution{Digital Learning Hub.; Ministère de l'Éducation nationale, de l'Enfance et de la Jeunesse, Luxembourg}
  \city{Esch-sur-Alzette}
  \country{Luxembourg}}
\email{info@rohles.net}

\author{Yuwei Chuai}
\affiliation{%
  \institution{SnT,University of Luxembourg,}
  \city{Luxembourg}
  \country{Luxembourg}}
\email{yuwei.chuai@uni.lu}

\author{Gabriele Lenzini}
\affiliation{%
  \institution{SnT,University of Luxembourg,}
  \city{Luxembourg}
  \country{Luxembourg}}
\email{gabriele.lenzini@uni.lu}

\author{Anastasia Sergeeva}
\authornotemark[1]
\affiliation{%
  \institution{University of Luxembourg,}
  \city{Esch-sur-Alzette}
  \country{Luxembourg}}
\email{anastasia.sergeeva@uni.lu}

\renewcommand{\shortauthors}{Trovato and Tobin, et al.}

\begin{abstract}

Reducing the spread of misinformation is challenging. AI-based fact verification systems offer a promising solution by addressing the high costs and slow pace of traditional fact-checking. However, the problem of how to effectively communicate the results to users remains unsolved. Warning labels may seem an easy solution, but they fail to account for fuzzy misinformation that is not entirely fake. Additionally, users' limited attention spans and social media information should be taken into account while designing the presentation. The online experiment (n = 537) investigates the impact of sources and granularity on user’s perception of information veracity and the system’s usefulness and trustworthiness. Findings show that fine-grained indicators enhance nuanced opinions, information awareness, and the intention to use fact-checking systems. Source differences had minimal impact on opinions and perceptions, except for informativeness. Qualitative findings suggest the proposed indicators promote critical thinking. We discuss implications for designing concise, user-friendly AI fact-checking feedback.

\end{abstract}





\maketitle
\section{Introduction}

Social media platforms like Facebook, X (formerly Twitter), Reddit, WhatsApp, and TikTok have emerged as primary online locations for people to seek and share information. However, this has been accompanied by a significant increase in the spread of misinformation. Online misinformation includes (re)posts, comments, or statements that are misleading due to inaccurate, incorrect, or biased content, regardless of being part of any deliberate plan to deceive for a specific objective or propaganda.\footnote{This latter is called disinformation, see for instance, CISA's \url{https://www.cisa.gov/topics/election-security/foreign-influence-operations-and-disinformation}}Recent developments in Generative AI have increased the problem of online misinformation by making the large-scale creation of realistic fake content significantly easier \cite{edmoGenerativeMarks}.
To comply with regulations \footnote{ \url{https://www.cima.ned.org/publication/chilling-legislation/}} and provide better services to their users, social media platforms are trying to reduce the spread of misinformation.
To achieve this goal, they actively collaborate with independent fact-checker organizations \cite{facebookIntoFacebook,tiktokCombatingMisinformation,aghajari2023reviewing} who, using the practices of journalists' investigation, analyse the claims and provide a verdict. However, as was widely reported, fact checking is labor intensive and often not timely, with results frequently appearing days later. Furthermore, only a small portion of the content is fact-checked \cite{avaazFacebooksAlgorithm}. Therefore, some key platforms stopped these fact-check initiatives and began to explore alternative solutions to assess the credibility of the content. One of these is to rely on crowdsourcing credibility ratings such as X's "Community Notes". However, recent studies have shown some negative effects provided by the "Community Notes" on the emotional climate on the platforms \cite{chuai2024community}. 

Another alternative can be the larger exploration of automatic credibility assessment methods powered by Artificial Intelligence (AI). The implementation of AI-based fact-checking instruments can significantly increase the efficiency and coverage of the procedure. Modern AI-based methods can imitate the activities of human fact-checkers, including claim detection, evidence retrieval, verdict prediction, and justification procedures \cite{guo2022survey}. Moreover, they can create an opportunity to produce detailed (not just true/false) ratings of credibility using a big range of sources and attending weights to different parameters of information \cite{guo2022survey}. Several papers investigates such parameters of information for automated content verification, such as emotions \cite{liu2024emotion} or text coherence with existing ``approved'' knowledge bases like Wikipedia \cite{thorne2018fever}, but also extra-textual cues like source reputation \cite{pillai2024hierarchical} or social clues like likes and reposts (e.g., \cite{yang2019unsupervised} similarly to the strategies performed by fact-checkers \cite{juneja2022human} and users \cite{tang2024knows}).

While progress has been made in AI detection and verification, effectively communicating the results of AI-based fact-checking to users remains an open challenge \cite{guo2022survey}—this challenge requires rethinking the current approaches to interacting with users. Currently, social networks mainly use warning labels, which inform users about the fact-checking results \cite{lim2023effects}. Although research has shown that these measures inform users about misinformation \cite{martel2023misinformation,mende2024fighting}, several concerns could affect their efficiency. One concern is the complex nature of online misinformation, which often cannot be neatly categorized as either true or false \cite{martel2023misinformation}. 
This complexity can arise from conflicting evidence or from blending factual information with personal opinions and experiences in posts \cite{zhao2021detecting,li2022health,krause2022infodemic}. Yet existing labels frequently classify content with conflicting evidence as false, which oversimplifies the issue and provokes questions about their trustworthiness \cite{stomber2021towards}. Another issue is that current indicators give users little insight into why content is marked as false. For example, platforms use warning flags referencing third-party fact-checkers \cite{brandtzaeg2016emerging}.\footnote{Recently, Facebook announced the shift from that model towards a crowdsource-based one (https://www.nbcnews.com/tech/social-media/meta-ends-fact-checking-program-community-notes-x-rcna186468). However, at the moment of submission, these changes were not yet being implemented.
} However, these fact-checker findings are typically hosted on separate websites, not adjacent to the flagged material. 
This means users must take extra steps to find the information they need, making the process long and complicated \cite{tang2024knows}. 
Several studies showed that an explanation of the reasons near the warning labels could help users make a decision about information veracity \cite{epstein2022explanations,martel2023misinformation}. However, which type of information should be put inside these labels requires future exploration \cite{spradling2021protection}. 

To better communicate the results of the AI fact-checking system to users, these systems should avoid the problems listed above. In addition, they should take into account the context in which misinformation appears on social media. Due to the overload of information on social media, users often make quick decisions and do not always want to engage deeply with the content and debunking messages \cite{talwar2019people}. This means that explanation modes involving long and detailed text will probably not be applicable, leading to the need to present information in a condensed way that still provides users with enough content for reasoning and decision-making.

In the present paper, we investigate several ways to enhance the communication of post-veracity by understanding which information should be included in AI-based fact-checking and misinformation warning system presentations. We look into what works better in warning users and what information a future misinformation warning system should report.
We conducted a pre-registered, human-subject experiment with 537 participants, drawing upon literature on interface cues and decision support systems in misinformation evaluation. 
We simulated scenarios in which individuals read and assessed the veracity of information. We then varied the sources of credibility. Specifically, we distinguished content-based cues, like the writing style, or factual correctness of information according to scientific consensus (content-based indicators) and cues outside the content, like source reputation or other users' social reactions (context-based  indicators). We also varied the system's indication of veracity, presenting the probability of information being true or false rather than a binary true/false indication. This experiment explores how these conditions affect users' judgments of news credibility and their trust in AI-driven fact-checking aids.

We found that both content-based and context-based  credibility indicators significantly aid users in evaluating the truthfulness of information without notable differences in opinion changes, system competence, and trust levels, except for informativeness. This underscores the value of certain meta information in veracity judgments, especially in environments like encrypted messaging platforms where direct content assessment is not feasible. Furthermore, we found that fine-grained indicators can provide more nuanced opinions, empowering users to make informed decisions about the content they encounter. We also discovered that the granularity of these indicators and the certainty they convey significantly affect user experience and perceptions of information veracity, highlighting their potential in scenarios where information is not fully verified. Finally, we found the explanations constructed around these indicators enhance critical thinking, encouraging users to reassess information based on the credibility cues provided, thereby promoting a more critical engagement with content.

Our contributions are as follows:
\begin{itemize}
    \item We made a comprehensive examination of how a system's assessment of social media post veracity influences users' perceptions of its credibility. This analysis helps to understand the interaction between automated veracity checks and user trust.
    \item We provide an analysis of how the non-binary indicators of content veracity and sources of the system's fact-checking results shape users' opinions about information credibility.
    \item We demonstrate that systems informing users about the trustworthiness of information (beyond simply flagging false information) are perceived as valuable. Our research provides evidence of how such systems could assist users in developing a more nuanced approach to trusting online information.
    \item We discuss emerging trends in the realm of information veracity-checking systems and propose design recommendations. These recommendations aim to enhance the effectiveness and user-friendliness of future systems, contributing to a more informed and discerning online community.
\end{itemize}

\section{Related work}

The section is organized as follows:
First, we examine how current platforms convey credibility assessments through warning-based approaches and discuss why these methods often prove ineffective (see Section \ref{Warning Labels}). Second, we review AI-driven strategies to communicate the results of the credibility assessment and support more informed user decisions (see Section \ref{Displaying System Opinions}). Next, we consider how people process information and make credibility judgments, emphasizing the implications for designing the presentations of results of AI based credibility communication (see Section \ref{human process information}). Finally, we explore how users leverage various credibility signals when checking facts (see Section \ref{internal_sources} and \ref{context-based _sources}  ).

\subsection{Communicating Content Veracity through Warning Labels}
\label{Warning Labels}
Warnings, as a form of security-enhancing friction, play a vital role in mitigating risky behaviors in cybersecurity \cite{distler2020framework}. Social media platforms leverage misinformation warnings with security-enhancing friction to improve users' understanding of news accuracy and direct their attention to news credibility, thereby informing and guiding their actions. These warnings are divided into two types based on the level of friction: contextual warnings, which highlight misinformation without restricting content access, and interstitial warnings, which interrupt the user and require interaction \cite{wu2006security,egelman2008you,kaiser2021adapting}.
For example, X employs interstitial warnings by covering content and requiring user engagement to access it \cite{Roth_Pickles_2020}. In contrast, platforms like Facebook, Instagram, and TikTok use contextual warnings, such as stop icons and textual labels directly on or alongside content identified as false or misleading (e.g., disputed or false information) \cite{Lyons_2018,tiktokUpdateWork}. Currently, these labels not only inform users that the content is disputed but also link to further information about the dispute.

While warning labels can encourage more careful engagement with online content, reduce the impact of fast thinking, and decrease the perceived credibility of fake news and the intention to share false information \cite{van2018future,yaqub2020effects,oeldorf2020ineffectiveness,wintersieck2021message}, they still have some limitations. Users may overlook general contextual warnings due to habituation and the false alarm effect, which refers to the skepticism towards warnings caused by previous encounters with inaccurate alerts \cite{guo2023seeing}. Additionally, current warning labels often display only the final results of fact-checking, requiring users to click for more information, which sometimes leads to the platform’s general rules rather than specific explanations about why the content is false. This lengthy process can reduce the effectiveness of warning labels, as users often make quick decisions about information \cite{tang2024knows,kirchner2020countering}. Moreover, research shows that people tend to seek like-minded information and avoid content that conflicts with their preexisting beliefs. Consequently, users may avoid clicking on warning explanations that contradict their beliefs, reducing the effectiveness of current warnings in correcting misperceptions and promoting analytical thinking \cite{tanaka2023does}.

To address this challenge, Guo's research emphasizes that specific contextual warnings, particularly with bright colors and explicit wording, attracted more attention and engagement compared to general contextual warnings \cite{guo2023seeing}. Additionally, studies show that providing a clear and comprehensible reason or context for classifying a post as misinformation significantly influences user engagement and makes the warnings more noticeable and effective \cite{lim2023effects,sharevski2022meaningful}. Kirchner and Reuter's study further suggests that adding an explanation to warning messages is the most effective warning-based approach \cite{kirchner2020countering}.

\subsection{Displaying System Opinions in Misinformation Correction Tasks}
\label{Displaying System Opinions}
In the context of combating misinformation, several empirical studies have been dedicated to understanding how to present the results of AI-system veracity evaluations to users in a way that helps them assess the truthfulness of the information. Horne et al. demonstrated the efficiency of different modes of AI-assistant explanations (including a fine-grained indication of pro and contra evidence) in misinformation detection tasks \cite{horne2019rating}.
The study by Lai and Tan indicates that machine learning assistance, particularly through explanations and predicted labels, improves decision-making in deception detection tasks \cite{lai2019human}. 
Lu et al. explored the effects of AI-based credibility indicators on different stages of the misinformation-detecting process and their willingness to share news under social influence 
 \cite{lu2022effects}. 
In addition, Si and colleagues investigated the use of Large Language Models to assess the truthfulness of claims, providing textual explanations to aid users distinguish between true and false information \cite{si2023large}.
At the same time, studies have shown some drawbacks to user trust and performance when the model explanation presentation does not align with the user's mental model of the news \cite{mohseni2021machine}. 
In the same line, the findings of a meta-analysis showed that length and lexical complexity are rather negative predictors of efficiency in the context of misinformation debunking \cite{walter2020fact}.
Previous research demonstrated that implementing a system's assessment of credibility through \%-based indicators of truthfulness notably enhances users' trust in the system and aids in their comprehension of its decisions \cite{chien2022xflag}. However, the specific relationships between the proposed metrics and the degree of change in users' opinions remain unestablished. Similarly, the study of Horne et al. \cite{horne2019rating} showed the general positive effect of the explanations on users' reliability perception, but the role of the exact indicator of trustworthiness remained unclear.

\subsection{Cognitive Processing and Evaluation of Information Credibility}
\label{human process information}
From a cognitive perspective, users process information and misinformation through the same mechanisms \cite{marie2020cognitive}. Given the vast amount of information available online, individuals often experience information overload. Thus, they rely on heuristic processing and cues to determine which content is credible and worth their further attention. Extensive literature explains how people understand and evaluate persuasive arguments, including the Elaboration Likelihood Model \cite{petty2011elaboration}, the Heuristic-Systematic Model (HSM) \cite{chaiken1980heuristic,chaiken2014heuristic}, and the dual-process theory of Intuition (System 1) and Reasoning (System 2) \cite{kahneman2003maps}. In essence, two primary styles of information processing guide these judgments: heuristic and systematic \cite{chaiken1980heuristic,chaiken2014heuristic}.
Heuristic processing relies on noticeable and easily comprehended cues as shortcuts for opinion formation. Systematic processing involves thorough understanding of all available information through careful elaboration and slow thinking.
Research indicates that systematic elaboration is associated with a lower susceptibility to fake news \cite{pennycook2021psychology}. However, systematic processing requires mental effort, and people engage in it only when they are able and motivated to invest the necessary resources \cite{chaiken2012theory}. Due to information overload, users often lack the cognitive capacity or time to evaluate information systematically and instead use heuristic processing to assess credibility. As a result, quick, heuristic-driven judgments are often made before more systematic cognitive analysis of message credibility \cite{chen1999heuristic,metzger2013credibility,metzger2010social}.


Message credibility can be defined as \textit{“an individual's judgment of the veracity of the content of communication”}, driven by perceptions of authenticity, accuracy, and believability \cite{appelman2016measuring}. 
 Lucassen and Schraagen’s (2011) 3S-model \cite{lucassen2011factual} conceptualizes this evaluation through three strategies: (1) examining semantic features within the content itself (e.g., factual correctness, clarity); (2) assessing surface features (e.g., layout, images); and (3) drawing on prior experiences with the source. The first two strategies often require active, systematic scrutiny, while the third relies on heuristic cues rooted in familiarity and trust of source. Other studies further suggest that credibility assessment depends on factors such as contextual cues, and consistency with prior knowledge or other resources\cite{kuutila2024revealing,savolainen2023assessing,lin2016social}. Given that systematic, in-depth analysis demands motivation and domain expertise, resources that can be scarce in a high-volume information environment, most users ultimately rely on heuristics. To clarify how these heuristic judgments are formed, it is useful to distinguish between factors that are intrinsic to the content and those that provide context-based  context for interpretation. For example, Savolainen divided the factors into two groups: related to the message (text-related factors) and to the author of the message (source-related factors) \cite{savolainen2023assessing}. In line with this approach, we divide credibility factors into two broad groups: Content-based, capturing intrinsic attributes of the message (e.g., factual accuracy and linguistic style), and Context-based, encompassing contextual elements such as source credibility and social endorsements.


\subsection{Content-based Credibility Assessment}
\label{internal_sources}
\subsubsection{Content Verification through Facts Verification}
In the past, fact-checking was a practice primarily linked to professional journalism, as a part of a journalist's expertise and part of pre-publication verification \cite{graves2019fact}. However, in the age of social media, fact-checking is done by non-journalistic organizations who act as external fact-checkers and verify already published information \cite{graves2019fact}. 
These organizations now encompass a wider range of participants, many of whom do not necessarily have a professional background in journalism \cite{cotter2022fact}. 
More recently, crowd-sourced fact-checking strategies based on non-expert users, such as Community Notes on X (formerly Twitter), provide a new way to limit the spread of misinformation and inform other users \cite{allen2021scaling,chuai2023roll}.
Studies provide mixed evidence about the effect of marking content as ``fact-checked'' on users' perception of the information \cite{walter2020fact}. In situations involving a high level of partisanship, indicating that the information has been fact-checked brings little or adverse effect \cite{jennings2023asymmetric,walter2020fact}.

\subsubsection{Characteristics of Writing Style}
Several works have focused on assessing credibility through content analysis, examining writing style, linguistic patterns, and semantic factors. For example, Horne and Adali analyzed the writing style of fake news and real news. They discovered that fake news often features longer titles, simpler words, and a higher frequency of proper nouns and verb phrases. In contrast, real news tends to have shorter and less repetitive content \cite{horne2017just}.
Research also showed that users adopt style-based heuristics to make suggestions about content veracity; for example, König et al. investigated the impact of different language styles (enthusiastic vs. neutral) on online health forums and found that the enthusiastic language style is perceived as less trustworthy and credible \cite{konig2019influence}.

\subsubsection{Emotion of the Content}

Emotions can be contagious to other social media users and are considered a main driver for information sharing \cite{goldenberg2020digital}. In this regard, misinformation is more likely to express anger on social media compared to true information \cite{prollochs2021emotions}. This phenomenon remained significant within the context of the COVID-19 pandemic \cite{chuai2022really}. Consequently, several studies consider emotion features in AI models to improve the performance of misinformation detection \cite{zhang2021mining}. Users actively employ emotional characteristics as a heuristic in assessing the credibility of information. Specifically, when the information contains emotion-provoking elements, non-partisan users are more inclined to use these emotional responses as a guiding principle, leading them to perceive the information as less credible \cite{ali2022effects}.

\subsection{Assessing Credibility Beyond the Content of Information (Context-based Sources of Credibility)}
\label{context-based _sources}
\subsubsection{Source Reputation and History of Information Endorser}
A significant category of cues is used in assessing the credibility of a source based on contextual details supporting the post. In the context of social networks, this includes evaluating the reputation of the source, such as the credibility of the source and the historical characteristics of the account \cite{im2020synthesized}. Previous studies showed the general moderating effect of source-perceived credibility on content verification in the health information domain \cite{avery2010role,bates2006effect}; 
Li’s work highlights the importance of authors in determining health claims on social media’s credibility \cite{li2023assessing}. Meinert and Krämer conducted a study examining how expertise, likes, shares, pictures, and user involvement influence the perceived credibility of politicians' posts on Facebook and found that source is a key factor of perceived credibility \cite{meinert2020cues}. 

Consequently, several platforms use the reputation of the source to combat misinformation. For example, YouTube has introduced source information panels in videos to help viewers discern content from trustworthy sources \cite{Graham_2021}. Previous works discussed providing provenance warnings when the source of information is unverified \cite{folkvord2022effect,sherman2021designing}. Feng’s work further explored how provenance information altered users’ accuracy perceptions and trust in visual content shared on social media. They found that providing provenance information often decreased trust and raised doubts about deceptive media, especially when it revealed manipulation \cite{feng2023examining}. Additionally, Zade et al. designed an intervention featuring a `tweet trajectory' to illustrate how information reached a user. This intervention provides `contextual cues,' including details about a Twitter account’s recent online activity, aiding in assessing the account's credibility \cite{zade2023tweet}.

\subsubsection{Bandwagon Characteristics of Information}
Social cues such as the number of views, likes, comments, and shares are also integral for evaluating credibility.
These cues, commonly referred to as ``bandwagon cues,'' trigger a credibility assessment based on the logic that \textit{`if many others approve of a story, it is likely credible'} \cite{sundar2008main}. Researchers point out that bandwagon heuristics is a powerful cognitive shortcut for evaluating online news \cite{sundar2007news}. This aligns with the growing evidence on the impact of bandwagon cues in shaping credibility perceptions.  
For example, Luo et al. evaluated the effects of Facebook likes on perceptions of message credibility and detection accuracy. Their findings indicate that a high number of likes increased credibility \cite{luo2022credibility1}. Similarly, the presence of likes and retweets on Twitter has been shown to increase users' belief that the tweet contains credible information \cite{aigner2017manipulating}. Moreover, Jin and Phua's research demonstrated that celebrities with a high number of followers are perceived as more trustworthy, leading to increased ratings of source credibility \cite{jin2014following}. Other studies also demonstrated that posts with a high number of likes and shares are perceived as more credible \cite{li2023assessing}. 
Metzger et al.'s research indicates that people often trust information and sources that are trusted by others, whether they are known individuals or based on extensive reviews from unknown people, typically without critically evaluating the content or the credibility of the source \cite{metzger2013credibility}; a recent study of Panizza et al. \cite{panizza2023online} confirmed that people usually take these metrics into account making their assessment.

\subsection{Research Questions}
\label{Research Questions}
Prior research underscores the imperative to enhance our understanding of how to assist users in assessing the veracity of online information, thereby facilitating informed decision-making. Current systems predominantly employ credibility labels that mark information as ``false,'' but previous studies indicate that users perceive this approach as condescending and lacking transparency \cite{stomber2021towards, kirchner2020countering}. This insight calls for a more nuanced approach in designing human-centered credibility communicated tools that respect diverse interpretations and emerging message encrypting systems.

Users, professionals, and crowdsourced fact-checkers evaluate the credibility of information by considering various aspects of truthfulness; some of these methods can be implemented into AI-based fact-checking tools. Most methods discussed in the modern literature are based on domain knowledge from existing verified sources or databases, and verify facts in social media posts against these sources. It has also been shown that extra-content information can aid in verification. 
While these features can be incorporated into AI-based fact-checking systems, it is not yet clear how to communicate to the user the information about the sources of AI-based assessment \cite{guo2022survey}. Several studies examine separate credibility cues\cite{liu2024emotion,thorne2018fever,pillai2024hierarchical,yang2019unsupervised}; however, to the best of our knowledge, no previous studies have compared user reactions to context-based  versus content-based sources of credibility explanations in misinformation debunking systems.
Also, despite the frequent discussion in the literature about the acknowledgement of uncertainty in 
misinformation debunking systems \cite{horne2019rating,chien2022xflag}, as a means of assisting users in forming opinions about the reliability of the information, there is a lack of research on how these approaches affect users' trust calibration to the information in the misinformation debunking setting.
 

To fill these research gaps, our study aimed to answer the following research questions (RQs): 

\textbf{RQ1}: How does information about the credibility of the information impact users' beliefs about it?

\textbf{RQ2}: Which has a greater influence on users' beliefs regarding the system quality and veracity of a topic: context-based  or content-based indicators of information credibility?


\textbf{RQ3}: Does the information provided by using fine-grained indicator topics veracity affect the user’s perceived beliefs about the provided information and the perceived quality of the system more or less than information provided using binary (either true or false) indicators?


\textbf{RQ4}: How does the degree of certainty regarding the truthfulness/falseness provided by the system affect users' perceptions of the topic's veracity?


\section{Methodology}
As outlined in section \ref{Research Questions}, this study focuses on the presentation of AI-based veracity assessments and their impact on users, not on the technical implementation of such an AI assistant. Following Human-Centered Design \cite{Human-Centered-Design}, we created a simulation of the AI assistant interface, allowing us to investigate our research questions in a controlled experiment. We will first explain the experimental materials and then outline the study procedures and measurements.
\subsection{Experimental Material}
\subsubsection{Posts for Veracity Assessment}
We followed the study by Panizza et al. \cite{panizza2023online} and chose health-related scientific content for our materials. We chose health-related topics over alternatives (e.g., politics) because, with some exceptions like vaccination, they do not draw particular public attention. Therefore, we expected a lower controversy or partisanship. 
Drawing on medical blogs and sites about health misconceptions,\footnote{e.g., \url{https://valleywisehealth.org/blog/medical-myths-debunked-truth-behind-common-health-misconceptions/}
\url{https://www.gleneagles.com.sg/health-plus/article/10-health-myths-debunked}
\url{https://www.webmd.com/balance/ss/slideshow-10-health-myths-debunked}
\url{https://my.clevelandclinic.org/health/diseases/10350-peptic-ulcer-disease}
} we selected three false and three scientifically supported claims. Based on these claims, we employed ChatGPT-4 to create six posts for a fictional social network community promoting health and lifestyle. We purposely opted for text-based posts rather than image- or video-based posts, similar to platforms like X or Facebook rather than Instagram or TikTok (see Fig. \ref{fig:Example Interface} for an example). We took this decision based on the prominence of these networks and on the ease of sharing content in a text-based form. To ensure accuracy and minimize bias, the first and last authors carefully reviewed and revised each post, aligning them with verified medical blogs and sites. We then conducted a pilot test to refine the posts' fluency, style, and realism to mirror typical social media content. The generated posts are presented in Appendix.\footnote{
We decided to implement artificial posts instead of real ones because of ethical concerns related to privacy and the potential emotional impact on individuals whose content might be analyzed without their consent.}
\subsubsection{Fine-Grained and Binary Indicators}
The ``Binary'' indicator displayed the word ``true'' or ``false,'' accompanied by the text: ``The system predicts that the claim in this post is [true/false].'' This indicator was always aligned with the scientific consensus.

The ``Fine-Grained'' indicator presented a pie chart showing the proportion of evidence supporting the post as being ``true'' and ``false.'' The accompanying text read: ``The system predicts a [X]\% probability that the claim in this post is true and a [100 - X]\% probability that it is false.'' The exact value of X was generated as a random number ranging from 60 to 95 \% if the original claim was supported by scientific evidence (we call these posts ``rather true'' in the paper) and from 40\% to 5\% if it was not supported (``rather false'' in the paper). We anticipated that this range would allow to test how differences in certainty would affect the magnitude of change in the user's opinion.
\subsubsection{Context-based  and Content-based Sources of Credibility}
\begin{table}[]
\caption{Types of Indicators and Heuristics, used in the study}
\label{table:Kinds of Indicators and Heuristics}
\renewcommand\arraystretch{2}
\resizebox{\columnwidth}{!}{%
\begin{tabular}{lll}
\hline
Type of Indicator  & Heuristic           & Presentation in study                                                                                \\ \hline
Content-based Indicator & Factual correctness \cite{walter2020fact,lederman2014can, zhi2017claimverif} & \makecell[l]{Verification of claims presented in the post \\against scientific databases.  }                         \\
                   & Writing style  \cite{konig2019influence,horne2017just,rathod2022exploring}     &\makecell[l]{ Presence or absence of linguistic patterns\\ and writing styles commonly found in misinformation.   }   \\
                   & Emotional content \cite{ali2022effects, prollochs2021emotions, zhang2021mining}  & \makecell[l]{Presence or absence of sentiment-related features\\ typically associated with misinformation. }         \\ \hline
Context-based  Indicator & Source reputation~\cite{meinert2020cues,Graham_2021,folkvord2022effect}   & Credibility of the source.                                                                           \\
                   & Source history   ~\cite{zade2023tweet,lin2016social,feng2023examining}   & \makecell[l]{Presence or absence of account activity patterns associated \\with the distribution of misinformation.} \\
                   & Bandwagon cues   ~\cite{sundar2008main,luo2022credibility1,li2023assessing,panizza2023online}   & \makecell[l]{User engagement indicators, such as the number of views, \\likes, and shares.   }                       \\ \hline
\end{tabular}%
}
\end{table}
We created ``Content-based'' and ``Context-based'' sources' credibility explanations based on the credibility assessment heuristics presented in the literature. Our inclusion criteria were: a) the heuristic was mentioned in multiple papers; b) evidence from previous literature suggested that this heuristic is understandable by users and can be used as part of their information verification process. We crafted and refined the explanations to ensure readability. The explanations used for two types of source credibility are presented in Table \ref{table:Kinds of Indicators and Heuristics}. 
Further modifications were made based on feedback from the pilot test. 
The interface of the AI-fact-checking system 
is presented in Fig. \ref{fig:Example Interface}.


\subsection{Experimental Design and Measurements}

\begin{figure}[h]
  \centering
  \includegraphics[width=1.1\textwidth]{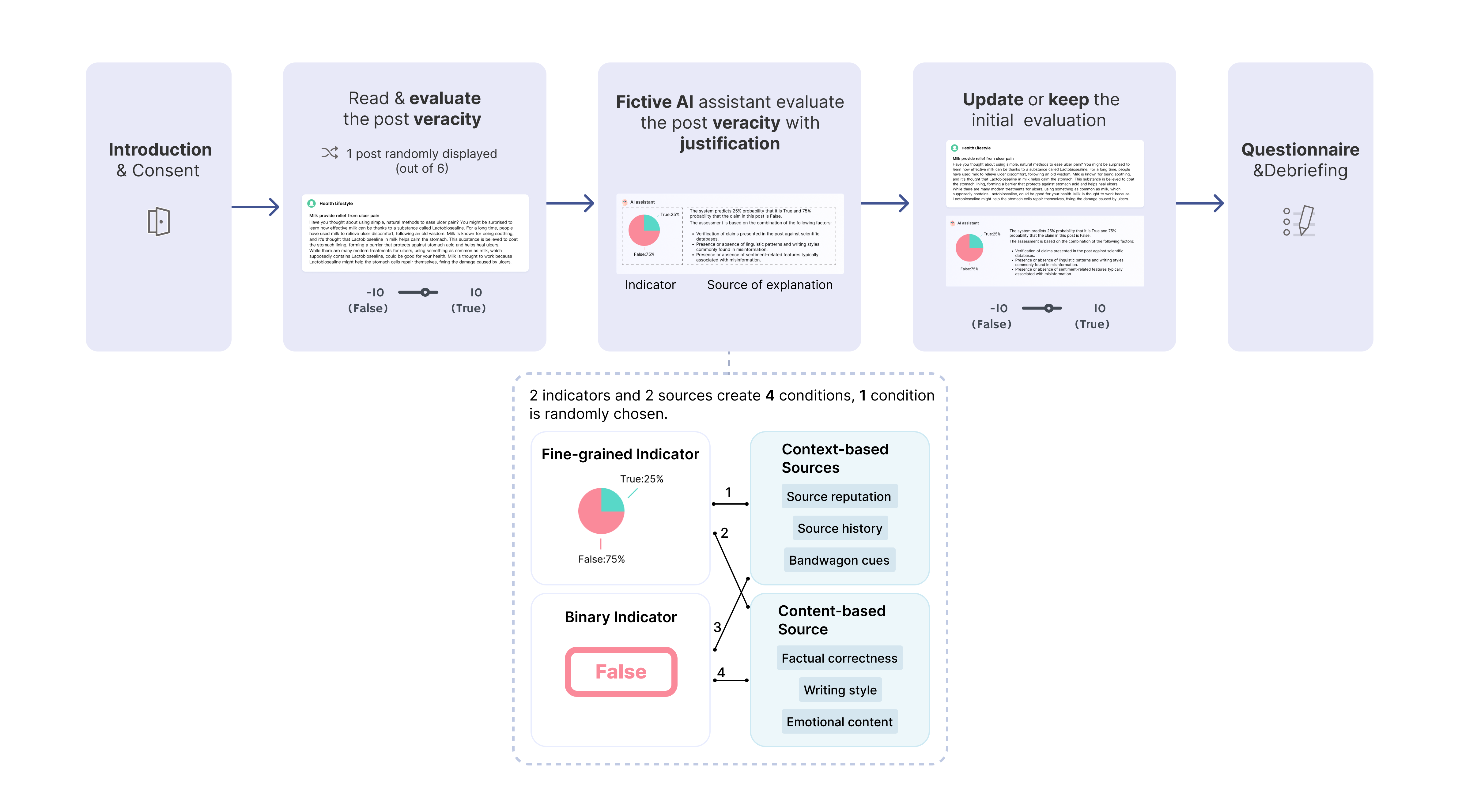}
  \caption{Experimental procedure}
  \label{fig:Experimental Procedure}
  \Description{The experimental procedure flow: (a) Introduction \& Consent: Introduce the experiment task and procedure. (b) Read \& Evaluate the Post Veracity: Participants read a post and make an initial judgment about its credibility. One post is randomly displayed from a set of six posts. (c) Fictive AI Assistant Evaluation: The AI assistant provides a credibility assessment result of the post with an explanation. Participants are presented with one of four randomly chosen AI-based credibility indicators, which include a fine-grained indicator or a binary indicator, each with either context-based  or content-based sources of explanation. (d) Update or Keep the Initial Evaluation: Based on the AI's assessment, participants have the option to update or keep their initial evaluation of the post’s veracity. (e) Questionnaire \& Debriefing: Participants answer open-ended questions and are debriefed about their exposure to misinformation.}
\end{figure}

The study employed a 2x2 factorial design to explore the impact of two independent variables: Source of Credibility in Explanation (Context-based /Content-based) and Indicators of Truthfulness (True or False or Fine-Grained). This design yielded four distinct experimental conditions (the experimental procedure is presented on the Fig. \ref{fig:Experimental Procedure}).

The study procedure was closely aligned with similar research on the influence of hypothetical AI systems on user behavior and perception, including the effects of presentation modes
\cite{lu2022effects,kim2024m,lee2023understanding,pareek2024effect}.
As measurements of the perceived quality of the system we used 
the set of heuristics, based on previous research, evaluating the quality of systems in recommender systems and conversational interface domains \cite{ribes2021trust,chien2022xflag,alarcon2023development,perrig2023trust,fogliato2022goes}.
We summarized them as a set of 7-point Likert Scales from ``Strongly Disagree'' to ``Strongly Agree'': 
 \begin{itemize}
     \item I think the AI assistant provided me with enough information to verify the truthfulness of the statement [INFORMATIVENESS];
     \item I think the AI assistant was useful in helping me verify the truthfulness of the statement [USEFULNESS];
     \item I trusted the information provided by the AI assistant [TRUST];
     \item I trusted the AI assistant's ability to help me verify the truthfulness of the statement [COMPETENCE];
     \item If the system is available, I would like to use it frequently [INTENTION TO USE].
 \end{itemize}

To minimize the risks of type I measurement errors, we restricted the set of measured heuristics to five, twice exceeding the "rule of thumb" of 50 participants per factor \cite{vanvoorhis2001statistical}
and preregistered all measurements.

As a control-for variable, we asked participants to rate their expertises in Health and Well-being and AI (7-point scale from 1, no knowledge, to 7, expert-level knowledge); we also collected information about gender, age and education level.

\begin{figure}
    \centering
    \includegraphics[width=\linewidth]{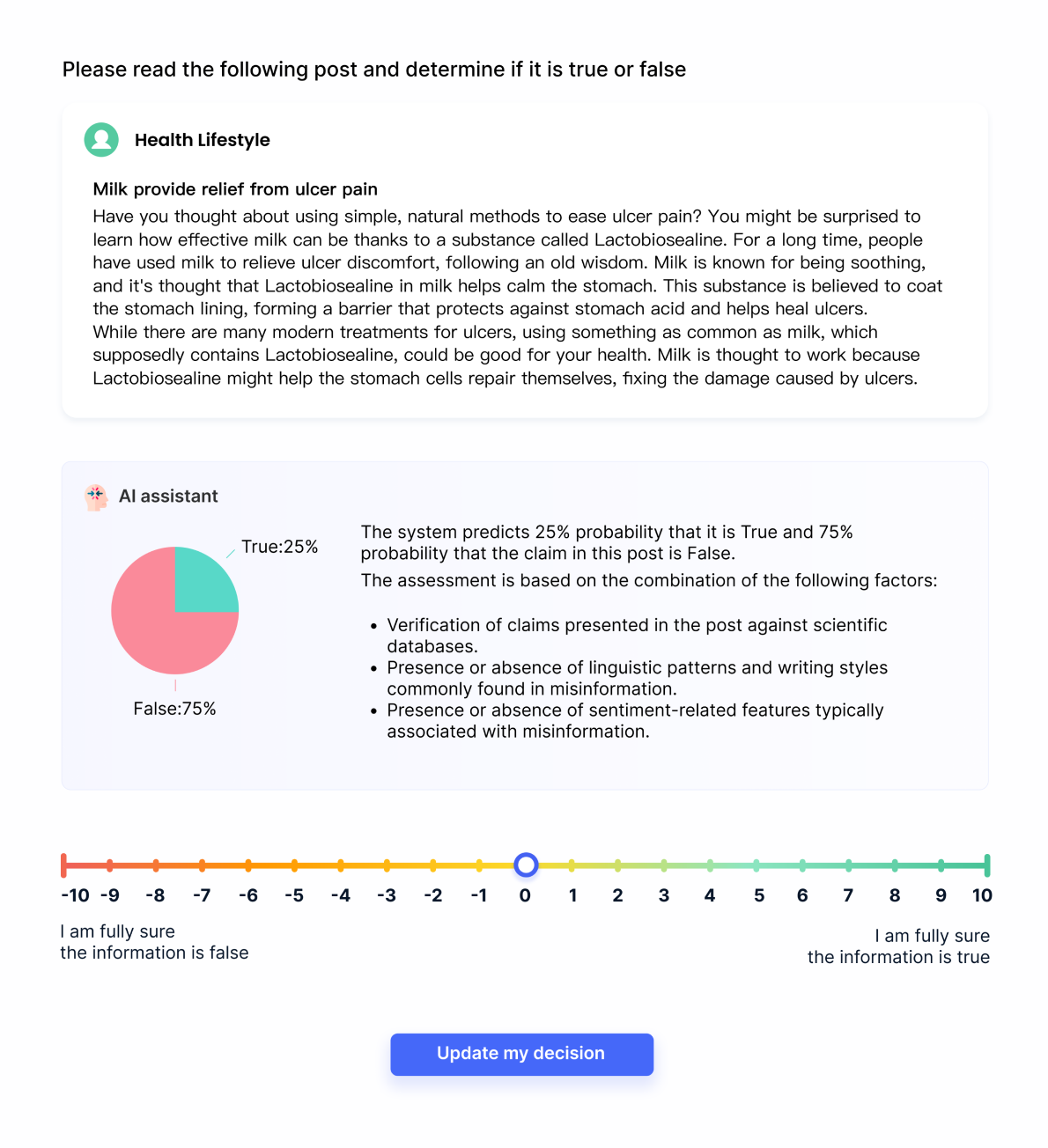}
    \caption{Example Interface Demonstrating the Integration of Fine-grained and content-based Indicators}
    \label{fig:Example Interface}
\end{figure}

\subsubsection{Procedure}
This study was preregistered on AsPredicted.org,\footnote{https://aspredicted.org/P18\_899} including information on our primary research questions, methodologies, and analysis plans. This study received ethical approval from the Ethics Review Panel of [anonymized]

Our target was to collect answers from around 550 individuals. We determined this sample size using a power analysis conducted in G*Power software, considering a small-to-moderate effect size reported in similar studies \cite{panizza2023online,lu2022effects,ribes2021trust} and the number of planned measurements.
Participants were recruited on Prolific. The study task was designed to take approximately 5 minutes with a compensation rate of 9 pounds per hour. Participants were screened based on the following criteria: residence in the US, balanced sample, English as a first language, a submission acceptance rate between 99\% and 100\%, join the platform not earlier than a year before the study was conducted. 

First, participants were briefed about the study procedure. They were informed that the study would evaluate health-related posts and warned not to participate if general health information could be sensitive to them. They were told that we are testing an assistant system (``AI-assistant'') that evaluates the veracity of the post.

After informed consent, participants were randomly assigned to one of the six posts. They were asked to evaluate the post's credibility on a scale from -10 (complete certainty that it is false) to +10 (complete certainty that it is true).

After their initial ratings, participants were presented with the AI assistant's credibility assessments and explanations. These assessments employed one of two indicator modes (``Fine-Grained'' or ``True or False'' indicators) with one of two explanations (``Context-based'' or ``Content-based''). Participants could use the AI assistant's clues or ignore them if they find the information untrustworthy or irrelevant. Upon seeing the AI assistant's insights, participants re-evaluated the veracity using the same -10 to +10 scale (the design of the reassessment page is presented in the Appendix).



Afterwards, participants rated their experience by responding to questions about the quality of the presentation.

 

Furthermore, participants provided open-ended answers on how they made decisions about the information before and after the AI assistant's intervention. With this qualitative data, we intend to gain insights into their decision-making rationale and their attitudes towards the AI assistant.

Finally, we disclosed the scientific consensus on the information, disclosed that the presented AI assistant does not yet exist as a working system and provided contacts of the researchers to further questions and to opportunity withdraw the consent.



\section{Results}
\subsection{Data Collection and Demographic}
We initially received a total of 560 responses. We first removed incomplete responses and calculated the mean completion time. In line with our preregistered criteria, we excluded responses with completion times less than two standard deviations below the mean (below 154 seconds). The resulting final dataset consisted of 537 responses.\footnote{After the publication, we will provide the full dataset on OSF platform} 268 participants identified as women and 269 as men. The average age of the participants was 39.4 (SD = 13) years. Educational background varied among the participants: less than 1\% did not complete high school, 11.5\% graduated from high school, 27.9\% had some college education, and 39.1\% held a Bachelor's degree. Additionally, 3\% had an unfinished Master's degree, and 17.6\% had a Master's degree and higher. The mean of expertise in AI was 3.61 (SD = 1.3), and in health and well-being was 4.36 (SD = 1.2). The data analysis was performed in SPSS 28 except for Equivalence testing, where we used the Python statmodels 0.14.1 package.

\subsection{Effect of Evaluation to the User's Perception of Post Veracity (RQ1)}
\subsubsection{General Effect of Advice on Opinion Change}
As a preliminary analysis, the Shapiro-Wilk Normality Tests showed a statistically significant departure from normality for the delta scores of opinion changes when the assistant suggested the post is true/rather true (W = .829, df = 258, p < .001) and false/rather false (W = .917, df = 279, p < .001). Therefore, we proceeded with Wilcoxon signed-rank tests. The tests revealed statistically significant shifts in the perceived veracity of posts influenced by the assistant's suggestions. This effect was observed regardless of whether the assistant indicated the post as true/rather true (Z = -12.294, p <.001) or false/rather false (Z = 9.973, p <.001), with changes aligning with the assistant's advice. 

Interestingly, when the AI assistant suggested that the information was true/rather true, we observed a 13\% binary opinion change (from thinking that the post is somehow ``false'' to somehow ``true''). In contrast, when the assistant indicated that the post was false/rather false, we observed a 56\% binary opinion change in agreement (from thinking that the post was somehow ``true'' to somehow ``false''). This difference appeared to be likely because the posts seemed believable to the users at first. So, when the assistant claimed a post was false/rather false, people were more likely to change their minds. This indicates how an AI assistant's advice can influence users' perceptions of post veracity.
The study observed that users generally changed their opinions following the assistant's evaluation. The magnitude of the change varied according to whether the evaluation was done. Specifically, for suggestions that a post was true/rather true, the average change in opinion was more modest (M = 1.76, SD = 2.65). In contrast, for suggestions that a post was false/rather false, the change was bigger (M = 3.36, SD = 3.6). This difference was not anticipated in our theoretical model and not included in our pre-registration. Therefore, we did not incorporate this finding into most of our analyses. However, we made an exception in the trust calibration analysis. In this case, we separately analyzed the data for positive and negative suggestions. This decision ensured our analysis remained robust and avoided potential biases such as Simpson's paradox \cite{chen2009regression}.
\subsubsection{Effect of Demographics and Level of Expertise on Opinion Change}
We ran a multiple regression model to determine the effects of demographic variables of age, gender, education level, and expertise in relevant domains (``Health'' and ``AI'') on the changes in opinion. The overall model did not explain a significant portion of the variance in 'DELTA' (Adjusted $R^{2}$ = .004, F(5, 522) = 1.437, p = .209);
\subsection{Effect of Context-based and Content-based Indicators of Veracity on User's Self-declared Trust and Opinion Change (RQ2)}
\begin{table}
\centering
\caption{Results of the Mann-Whitney U Test and TOST Test (Equivalence Interval -0.3 to 3) for Conditions with content-based and context-based Sources of Explanation}
\label{table:Results for content-based and context-based  Sources of Explanation}
\begin{tblr}{
  cell{1}{2} = {c=2}{},
  cell{1}{4} = {c=2}{},
  hline{1-2,9} = {-}{},
}
Parameters         & U-Mann-Whitney &               & Equivalence test
  (TOST)                     &                                            \\
                   & z-statistic    & p-value       & {t-statistic (lower;upper
\\~threshold test)} & {p-value (the biggest from\\~two
  tests)} \\
INFORMATIVENESS    & -2.194         & \textbf{.028} &                                               &                                            \\
USEFULNESS         & -.257          & .798          & 2.75;-1.95                                    & \textbf{.026}                              \\
TRUST              & -.293          & .769          & 2.58;-2.12                                    & \textbf{.017}                              \\
COMPETENCE         & -.586          & .558          & 2.91;-1.83                                    & \textbf{.033}                              \\
INTENTION
  TO USE & -571           & .568          & 2.72;-1.61                                    & .054                                       \\
DELTA              & .877           & .381          & 1.06;-1.056                                   & .146                                       
\end{tblr}
\end{table}

\subsubsection{Self-reported Metrics of Trust and Informativeness of the Indicators}
We ran U-Mann-Whitney tests to measure the effect of the context-based  and content-based indicators on informativeness, usefulness, trust, competence, and intention to use.\footnote{Contrary to what we had outlined in our preregistration, we did not conduct a comparison between all four groups via ANOVA model. This decision was made because the potential interaction effect between the source and the granularity of veracity was beyond the scope of our research questions.} We found that participants felt significantly less informed in the ``context-based '' source of explanation condition than in the ``Content-based'' condition (Z = -2.196, p = .028). 
However, we did not find significant differences between conditions in other indicators.

Following our preregistration, we ran the TOST Equivalence test \cite{lakens2018equivalence} with control for equality of variances by Levene's test. Based on the average of effect sizes estimates from \cite{ribes2021trust,lu2022effects}, which varied around small-to-moderate effect size, we considered our smallest effect size of interest equal to Cohen d =.3 (standardized equivalence boundaries -.3;3).
Our findings showed some evidence of the absence of significant differences between the groups by the parameters of the usefulness of the information, trust in the system itself, and the provided information. The findings are presented in Table \ref{table:Results for content-based and context-based  Sources of Explanation}. We calculated the median results for each parameter and found them equal to 5 (``rather agree'') from 7 points.
\subsubsection{Effect of Context-based  and Content-based Indicators on the Opinion Change}
\begin{figure}[h]
  \centering
  \includegraphics[width=1\textwidth]{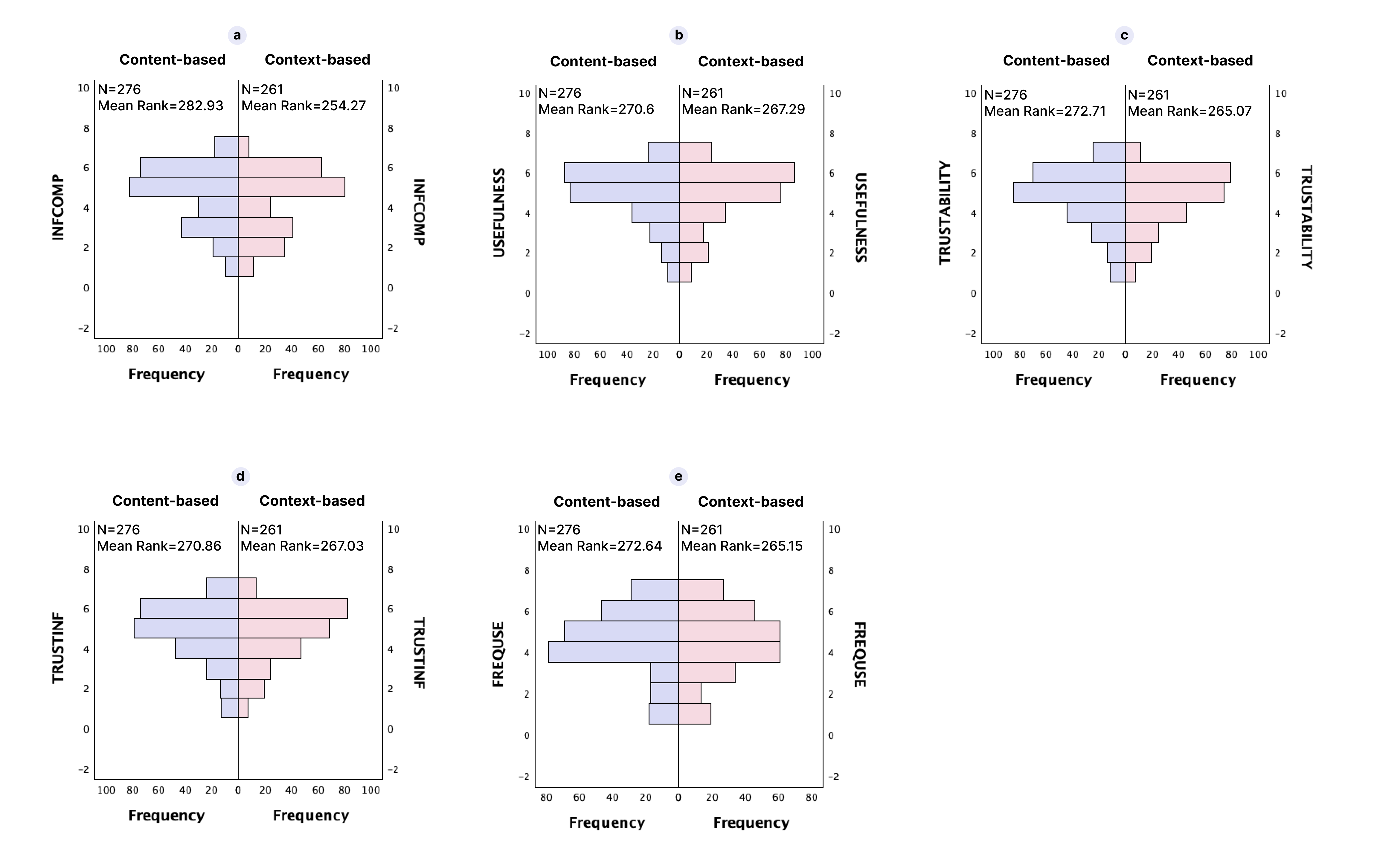}
  \caption{Results of Users Ratings of Content-based vs. Context-based  Credibility Indicators: Informative Competence (INFCOMP), Usefulness, Trustability, Trust in Information (TRUSTINF), and Frequency of Use (FREQUSE)}
  \Description{The results of user ratings of content-based and context-based  credibility indicators across five parameters: Informative Competence (INFCOMP), Usefulness, Trustability, Trust in Information (TRUSTINF), and Frequency of Use (FREQUSE). Each chart displays the distribution of ratings for content-based (purple) and context-based  (pink) indicators, along with mean ranks and sample sizes. The Mann-Whitney U test results show a significant difference for Informative Competence, with content-based indicators rated higher (p = .028). For Usefulness, Trustability, Trust in Information, and Frequency of Use, no significant differences are observed between content-based and context-based  indicators.}
  \label{fig:Internalcontext-based }
\end{figure}
To determine if the conditions affect the size of opinion change between the groups, we ran U-Mann-Whitney tests to the delta results of opinion changes. We did not find a statistically significant effect of the type of condition on the opinion change (Z = .877, p = .381). Following the TOST procedure with the same smallest effect size of interest considerations, we did not find significant evidence to confirm the absence of differences in the equivalence diapason (Table \ref{table:Results for content-based and context-based  Sources of Explanation}).

\subsection{Effect of the Fine-grained and True or False Indicators on the Parameters of Informativeness, Usefulness and User Trust (RQ3)}
\begin{figure}[h]
  \centering
  \includegraphics[width=1\textwidth]{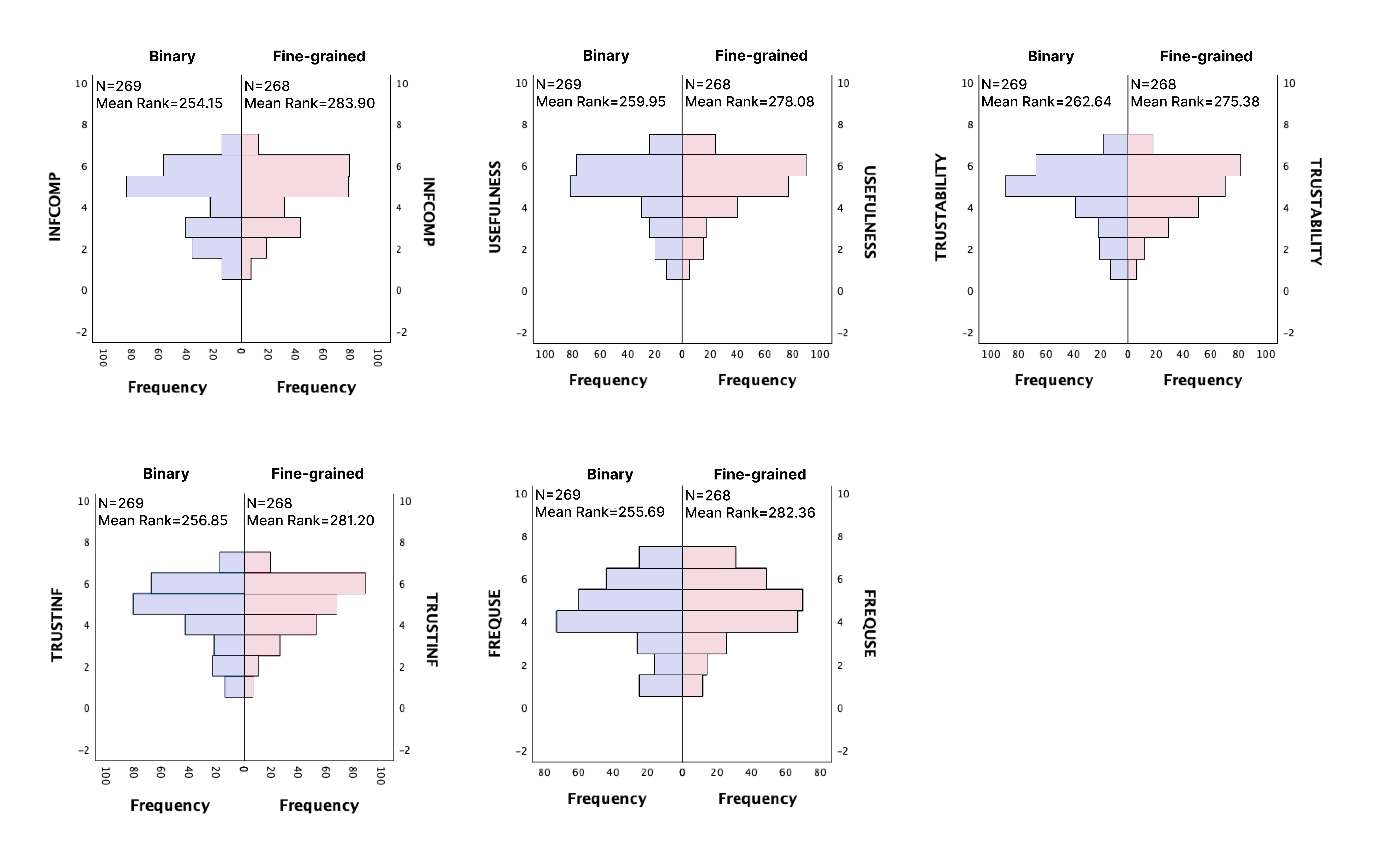}
  \caption{Results of Users Ratings of Binary and Fine-Grained Indicators: Informative Competence (INFCOMP), Usefulness, Trustability, Trust in Information (TRUSTINF), and Frequency of Use (FREQUSE)}
  \Description{The results of user ratings of binary and fine-grained Indicators indicators across five parameters: Informative Competence (INFCOMP), Usefulness, Trustability, Trust in Information (TRUSTINF), and Frequency of Use (FREQUSE). Each chart displays the distribution of ratings for binary (purple) and fine-grained (pink) indicators, along with mean ranks and sample sizes. The Mann-Whitney U test results show significant differences in Informative Competence and Intention to Use, with fine-grained indicators rated higher (p = .023 and p = .042, respectively). For Usefulness, Trustability, and Trust in Information, no significant differences are observed between binary and fine-grained indicators.}
  \label{fig:Internalcontext-based }
\end{figure}
\begin{table}
\centering
\caption{Results of the Mann-Whitney U Test and TOST Test (Equivalence Interval -0.3 to 3) for Conditions with Binary and Fine-Grained Indicators}
\label{table:Results for Binary and Fine-Grained Indicators}
\begin{tblr}{
  cell{1}{2} = {c=2}{},
  cell{1}{4} = {c=2}{},
  hline{1-2} = {-}{},
}
Parameters         & U-Mann-Whitney &               & Equivalence test (TOST)                      &                                         \\
                   & z-statistic    & p-value       & {t-statistic (lower;upper~\\threshold test)} & {p-value (the biggest from\\two tests)} \\
INFORMATIVENESS    & 2.279          & .\textbf{023} &                                              &                                         \\
USEFULNESS         & 1.396          & .163          & 0.60;-4.10                                   & .273                                    \\
TRUST              & 1.868          & .062          & 0.14;-4.58                                   & .443                                    \\
COMPETENCE         & .978           & .328          & 1.09;-3.67                                   & .139                                    \\
INTENTION
  TO USE & 2.032          & \textbf{.042} &                                              &                                         \\
DELTA              & -1.457         & .145          & 1.73;-0.4                                    & .345                                    
\end{tblr}
\end{table}
We conducted a U-Mann-Whitney test to compare the effect of the context-based /content-based indicators on informativeness, usefulness, trust, competence, and intention to use. We found significant differences between the groups by the parameters of informativeness (Z = 2.279, p = .023) and intention to use the system (Z = 2.032, p = .042). 
The TOST procedure did not provide evidence that the possible differences is smaller than the predefined diapason of equivalence (Table \ref{table:Results for content-based and context-based  Sources of Explanation}).
\subsection{Effect of the Fine-grained and True or False Indicators on the Opinion Change (RQ3)}
We did not find significant differences between groups on opinion change (Z = -1,457, p = .145). The TOST procedure did not provide evidence that the possible differences are smaller than the predefined diapason of equivalence (Table \ref{table:Results for Binary and Fine-Grained Indicators}). 

\subsection{Effect of Fine-grained Indicators of Content Veracity on the Magnitude of Opinion Change}
To determine the effect of fine-grained indicators on users' trust, we separately examined the relation between delta changes and the probability of the information being true in the condition of fine-grained indicators. Based on the results detailed in (the linked section), we conducted separate analyses for two distinct groups of questions (AI assistant suggesting that content is ``rather true'' and ``rather false'', respectively). Contrary to our preregistration, we opted for regression analysis instead of correlation analysis.\footnote{The reason for this change is that the correlation model is not suitable to account for the varying starting points of users before the system proposed additional explanations. For example, one person might initially choose 1, while another might choose 7, leading to different degrees of freedom to change their opinion even if equally convinced by the system.} We implemented two regression models, with the change delta as the dependent variable, the percentage of the fact being true, and the previous decision about veracity as independent variables separately to the case when the AI assistant suggested the post is ``rather true'' and when it suggested the post is ``rather false.'' 
In the ``rather false'' case, the model was significant (F(2, 276) = 22.296, p < .001, ( $R^{2}$) = .139). The initial score significantly predicted the delta score, $\beta$ = .375, t = 6.672, p < .001, while the shown probability did not ($\beta$ = .024, t = .436, p = .663). In the ``rather true'' case, both the initial score and the shown probability significantly predicted the delta score, F(2, 255) = 27.284, p < .001, with an $R^{2}$ = .176. The initial score was a significant negative predictor of the delta score, $\beta$ = -.387, t = -6.811, p < .001, while the shown probability was a positive predictor of the delta score, $\beta$ = .156, t = 2.745, p = .006.
\subsection{Qualitative Results}
After the study, participants answered open-ended questions about how they assess the post veracity, both independently and when assisted by AI. To understand their decision-making rationale and their attitudes toward AI assistants, we employed an inductive coding method  \cite{thomas2003general}. Initially, the first author analyzed 100 responses using open coding to identify emergent themes and establish a consistent coding framework. This framework was subsequently applied to all responses in MAXQDA software. The last author reviewed any discrepancies in coding to ensure consistency. We organized the responses into four principal themes: the influence of AI assistant on participants’ beliefs, participants' perception of source (content-based versus context-based  factors), participants' perception of indicator types (fine-grained versus binary) and the perceived influence of AI assistant's certainty on participant belief.

\textbf{AI Assistant's Influence on Participant's Beliefs}: Existing literature suggests that when evaluating the credibility of information, individuals often assess its consistency with their prior knowledge \cite{kuutila2024revealing,savolainen2023assessing,lin2016social}. Our study confirmed this finding. Approximately 54\% of participants mentioned that they relied on prior knowledge and personal experiences for initial judgments, seeking confirmation from an AI assistant: \textit{``The listing of transitory issues that aligned with my experience. The AI assistant helped to confirm my opinion.''} When the AI assistant aligned with their opinions, participants reported increased confidence in their judgments. Conversely, if the AI assistant's assessments contradicted their beliefs, they scrutinized the explanations provided to determine their trustworthiness, leading to two distinct behaviors. A considerable group of participants (36\%) used common sense and intuition to evaluate credibility. They found the AI’s explanations useful in decision-making, perceiving its insights and veracity assessments as logical and trustworthy: \textit{``The AI assistant showed good points as to how the post could be perceived as false, and since I had little knowledge to begin with, I trusted the AI''}.
In contrast, a smaller group (6.3\%) expressed skepticism towards the AI assistant's conclusions, citing a lack of sufficient information and evidence to support its judgment. They would not trust the AI assistant’s opinion: \textit{``The AI system was not helpful because despite it telling me what it found wrong with the information given, it did not give me resources to back that up. I can not blindly trust AI.''} 

\textbf{Perception of Source (Content-based versus Context-based )} Participants in both conditions found the AI assistant's explanations helpful and trustworthy, primarily because these explanations were consistent with their own reasoning and effectively demonstrated the AI's logic. This alignment fostered trust in the AI. While most participants considered the AI assistant’s explanations collectively, a few evaluated them independently. For instance, in context-based  indicators, participants identified historically spreading misinformation and source credibility as relevant for deciding on a posts' veracity, whereas bandwagon cues were deemed less relevant. One participant explained, \textit{``The AI assistant then said there was a large chance it was fake, and it gave me the reasons. The first reason was that it mentioned something like claims could not be validated. That is a huge red flag, so it made me even more certain that my initial thoughts were right- I hadn't heard of some of that stuff; the AI also said it couldn't validate it, so it must be false. The second reason AI gave (habits of having false information or something like that) was also a huge red flag. The third reason (likes, etc.) didn't really mean much to me.''} In the content-based condition, participants found all three credibility cues (consistency with scientific evidence, linguistic patterns, and sentiment features) useful and reasonable. Furthermore, across both conditions, participants reported that the AI assistant's explanations prompted them to reevaluate their initial assessments based on these credibility cues, leading to more deliberate decision-making: \textit{``The AI assistant made some compelling points about the falsehoods in the information that I hadn't considered initially, which influenced my final decision.''}

\textbf{Perception of Indicator Types (Fine-grained versus Binary)} We observed that presentation style influenced trust in the AI assistant. Participants generally preferred the detailed (fine-grained) mode, which increased trust: \textit{``With AI, the percentage estimate of how true the article was seemed legitimate and influenced my answer to change and believe the article as true.''} This aligns with prior research indicating that showing uncertainty can enhance user understanding and transparency in the AI's decision-making process, thereby fostering trust \cite{zhang2020effect,bansal2021does,marusich2023using,ma2023should,kim2024m}.
Additionally, users recognized that some topics do not have an absolute truth, and the fine-grained mode effectively captures this complexity. A participant noted, \textit{``The issue about eggs and health has been going back and forth for decades. While I thought the post was on the truthful side, I was not convinced it was totally true.''}. We also found that binary indicators could evoke distrust, as participants noted that this \textit{`all or nothing'} approach made them rely more on their prior knowledge.

\textbf{Perception of AI assistant's Uncertainty} Our study found that fine-grained indicators influence participants' judgments. Participants' scores tend to be aligned with the probabilistic true/false assessments provided by the system. For example, one participant remarked, \textit{``My original score and my amended score were very close to what the AI information had provided''}. We also noted that higher probability scores generally increased user certainty and confidence. However, when the probabilities were closely matched, the participants experienced confusion: \textit{ 'Since the AI assistant predicted 60\% as true and 40\% as false, it is almost like a 50\% 50\%. If the AI assistant prediction is more like 80\% 20\%, then I would be more influenced by those numbers.”} This indicates that while AI probabilities can influence user judgments, their effectiveness is diminished when probability values suggest ambiguity.

\section{Discussion}
\subsection{Veracity Checking System and Users Perception of Information Credibility}

In section \ref{human process information}, we outlined the need to balance communicating effective veracity information while avoiding increasing information overload. Prior studies emphasize the necessity of providing clear explanations for credibility assessments to enhance the effectiveness of AI-based credibility indicators \cite{epstein2022explanations,seo2019trust,horne2019rating}. Building on this work, our study explored presenting concise, understandable explanations of the AI system's verdict process. Unlike previous studies that presented true/false results with debunking narratives \cite{si2023large,hsu2023explanation,pareek2024effect}, our study demonstrated that even without specific debunking details, users could still gain valuable insights into information credibility from short indicators and the types of sources that contributed to the indicator. In light of the balance between communicating effective veracity information and avoiding information overload, these results are promising. 
The fact that the median results for scales of trustworthiness, trust in the information, and system usefulness were equal to 5 (``Rather Agree'') provides additional evidence that this presentation mode is well received by users. 
These rather positive results can be explained by the fact that the system's reasoning matched the users' typical reasoning (heuristics) to verify content. This might make them perceive the results of the automatic evaluation as reasonable and trustworthy.

Our results are in line with previous findings 
suggesting that shifting from warning labels to informative labels with feedback on information quality and potential risks could be beneficial for informing users about information issues \cite{konstantinou2024exploring, spradling2021protection}.

Noteworthy, although the interface provided only minimalistic reasons for its decisions, users value this for making decisions. This suggests that providing some explanations alongside warning labels might aid in forming an opinion about the information while avoiding increasing feelings of information overload. 

\subsection{Content-based and Context-based  Indicators of Credibility}
Our study examined the efficacy of credibility heuristics (content versus extra-content) to assess post veracity. Both types of indicators were instrumental in aiding users' decision-making processes. Moreover, we found that even concise indicators can promote systematic thinking. The qualitative feedback showed that providing explanations of these heuristics could prompt users to re-evaluate the authenticity of posts. These insights contribute to the ongoing exploration of how to effectively and understandibly communicate credibility assessment results.

For all of those credibility heuristics, participants expressed skepticism toward the effectiveness of bandwagon cues in determining credibility. This contradicts previous research suggesting that a high number of social endorsements increases perceived credibility \cite{luo2022credibility1,li2023assessing}, and aligns with recent research that shows bandwagon cues do not influence the evaluation of the trustworthiness of news \cite{koch2023effects}.
This discrepancy can be explained by the HSM model (See section \ref{human process information}): bandwagon cues are cognitive shortcuts for quick judgments. However, the AI assistant’s explanations are designed to encourage systematic processing. When individuals are motivated to understand and evaluate information thoroughly, they are more likely to engage in systematic processing, reducing reliance on heuristic cues like social endorsements. Furthermore, the community that provides social cues should be relevant to the person to have an impact \cite{savolainen2023assessing}. However, in our study, participants were merely informed that some other users liked and reposted the post, which might not provide enough information to decide whether their opinions should be taken into account.

We found that participants felt less informed in the ``Context-based '' source of explanation condition compared to the ``Content-based'' condition. 
In the ``Content-based'' condition, the AI assistant did not assess the content itself but relied on factors beyond the content, such as source reputation, source history, and contextual information like bandwagon cues. When users employed systematic processing, they felt that non-content cues did not provide enough information to determine content credibility, making them perceive the ``Content-based'' condition as less informative than the ``Context-based '' condition. However, we did not find significant differences between conditions in other parameters such as trust, intention to use, and opinion changes. These findings are in line with research by \cite{savolainen2023assessing} who showed that both reputational and message-based cues (e.g., correctness and logic of information) are important in credibility assessment on Reddit. Interestingly, as in our study, the presentation of AI worked as a proxy to mental assessment (participants evaluated the AI-system reasons rather than the original post), but this still affected their opinions. Potentially, the author's credibility (i.e., source) is an important heuristic for credibility assessments. If a source is promptly indicated as unreliable, it serves as a crucial heuristic to assist users in making decisions.

\subsection{Effect of Granularity of Opinion to the Users' Perception of Information Credibility}
Previous studies \cite{mohseni2021machine,chien2022xflag, horne2019rating} have shown that providing users with evidence of pro and contra of information veracity is beneficial for understanding the AI's decision. Our study contributes to this research by providing evidence that different levels of AI assistant certainty impact user certainty about misinformation. As it is often difficult to prove or fully debunk information, a fine-grained model can represent reality more honestly and help users build nuanced opinions. 

Moreover, presenting percentages of truthfulness or falseness aids users in deciding how to interact with the information and potentially avoid false alarm effects and habituation. Habituation effects refer to the \textit{``decreased response to repeated stimulation''} \cite{groves1970habituation}. This can cause users to pay less attention to warnings they have seen repeatedly and decrease their responsiveness over time \cite{guo2023seeing}. Therefore, it is essential not merely to convey information about content veracity but also to tailor warning mechanisms to better guide user behavior. One effective approach is the use of fine-grained indicators. These indicators enable users to make quick decisions without overburdening them with unnecessary details (as is currently implemented in standard warning systems). If the information's veracity is shown with high certainty, users can confidently accept or dismiss it. This streamlined decision-making process is efficient and minimally disruptive and does not influence users' main workflow on social media platforms. Conversely, in highly ambiguous situations, fine-grained indicators nudge users towards more systematic thinking. For example, if an indicator shows a balanced probability, such as 55\% true and 45\% false, it signals that the information is not straightforward. In the qualitative feedback, our participants reported cognitive unease with trustworthiness indicators close to 50\%. This ambiguity can make users doubt the AI assistant's quality. However, it might also prompt users to delve deeper, seek additional evidence, and engage with the information more thoroughly. Consequently, while our study highlights the value that such balanced probability markers could bring, future work needs to address precisely how the algorithm should work: Should it give the balanced probability for the post as a whole or for each individual statement made in the post? How would it handle the issue that not some statements are more central than others? How would it factor in links to further information that a post might include?

Interestingly, in our study, the effect of adjusted granularity worked only when the AI reported that information was likely true and not in the opposite case. This may be connected to the current users' mental model, which predominantly features binary indicators of true and false. Consequently, if information were marked as leaning towards being false, it was potentially perceived as simply false, without nuances. That might show the psychological difference between perceiving something as ``false'' and as ``true.'' We demonstrated that this effect could influence the use of fine-grained warning labels, which leads to the conclusion that systems should employ different visual and explanation patterns when suggesting that content is rather true or rather false.



\subsection{Design Implications}
\subsubsection{Customisation of Indicator's Information Support}
Our results showed that fine-grained indicators are preferred over binary indicators, aligning with previous research against oversimplification. At the same time, we agree with the cautious approach of Guo et al. \cite{guo2022survey}: Although fine-grained labels in AI-based fact-checking resolve the oversimplified true/false dichotomy, they are still not fully explanatory. For example, they do not provide much clarity when a claim consists of several sub claims, some of which are true and some false. We agree that it is necessary to split multiclaimed information into separate claims and process them individually \cite{guo2022survey}. Another proposed method involves text summarization approaches, where we first identify the main claim of the text and provide an indicator specifically dedicated to the core claim. By doing this, it is possible to add a one-sentence summary of the core claim of the post to the indicator while providing the metrics. Another important way AI-based indication can be implemented is by giving users a way to customize the type of information they prefer as the source of the rating (context-based , or different types of content-based features) and/or enabling them to switch between different sources to understand better why the claim appears problematic. As our participants mentioned, an indicator rating information as near 50/50 true-false creates confusion in interpretation. While we believe that a more accurate representation of information is positive, we also think the system should provide users with additional ways to overcome confusion and form their own opinions. For example, it could acknowledge uncertainty and suggest a set of links from both perspectives, allowing the person to make an informed decision if they wish.  
Previous studies also pointed to the fact that the status of the information can change over time \cite{tang2024knows}. As the current warning systems are not capable of capturing the changes in the information, we think that the future approach can incorporate the idea of updating the information in the informing labels and sending the updates to the users in case of severe changes (similar to the approach proposed by \cite{tang2024knows}; however, it should be implemented with cautions, as too many system notifications about information update can contribute to the information overload while using networks; as a possible solution it should be used only for the news, which is either marked as important by user or as such on the level of the whole network as the case of public safety (e.g. similar to the COVID crisis situation).

\subsubsection{Two-layers Verification Presentation}
Our results showed that even a short form of explanation provides the user with rationals for better decision-making.
However, these brief explanations seemed less effective when users had a strong pre-existing stance on the fact. In line with studies on system explanations in misinformation content, we recommend incorporating a second level of explanations into the design, such as links to external content or detailed system statistics. Such a two-level structure of AI recommendations on post veracity (fine-grained indicators on first sight, detailed information on clicking links) could help users reconsider their initial, heuristics-based evaluation, while also allowing them to switch to systematic processing of post veracity if need be, in line with HSM (see Section \ref{human process information}). In addition, we hypothesize that such detailed information might make it easy to understand how the percentage-based veracity indicators were created by the AI assistant, contributing to a deeper understanding of AI. It could also counter a potential misuse of our results: In principle, malicious actors could attempt to convine users of fake news by using purposefully manipulated fine-grained indicators (``This post is 77\% true'' rather than ``This post is true'', which might be too implausible). If users build the mental model of regularly checking both levels of veracity information, such malicious attempts would be more difficult to achieve.
\subsubsection{Design for Encrypted Social Media}
As our data showed, users favorably accept AI assistants based solely on context-based  indicators. While most current research discusses credibility indicators related to the content and form of messages, an emerging class of media makes their use impossible. For example, platforms like WhatsApp and Telegram use encryption protocols, meaning that only involved users can see the message content, making content-based analysis impossible.

However, multiple studies show that WhatsApp has become a major source of misinformation online \cite{madraki2021characterizing, martins2021covid19}, and users seek reliable ways to deal with this uncertainty. The current method implemented by the network (notifying users that content has been reposted many times) \cite{whatsappPreventSpread} is ambiguous, as this indication can be interpreted both as a warning and as a sign of content quality, creating confusion among users \cite{tang2024knows}. In this respect, indicators such as information provenance (source credibility at the community/person level, previously reported misinformation providers, or behavioral patterns like the speed of dissemination or reposting via specific communities) are possible metrics that can be used for rating formation and presentation of information to users. While content-based methods may still be preferred by users, the proposed approach can at least provide quick evaluation methods for these types of platforms.

\section{Limitations and Future Studies}
The study acknowledges the following limitations: 
The fact-checking process was performed and assessed through a single interaction with the indicators.
In actual use of the indication systems (e.g., Facebook warnings), users usually update their perception of system veracity via multiple interactions. Additionally, we divided the procedure into two steps to assess the opinion change effect; however, in the current indication systems, the warning typically arrived simultaneously with the main content (post). 
Consequently, the measurement approach might affect our findings regarding the system's perceived qualities. Although similar scales for measuring opinion changes \cite{panizza2023online,li2022health} have been utilized in previous studies, the scale measuring users' perceptions of information accuracy might not effectively capture changes in opinion. This is particularly concerning in cases where participants initially rate their certainty in the information's truthfulness as high, leaving no room to indicate an increased level of trust, should it occur (cellar effect).
The study focused on health-related (pseudo) scientific content, which may limit the results' applicability to misinformation from other domains. For instance, integrating source reputation and verification against domain knowledge may operate differently with political content; therefore, more studies are needed to determine the effect of partisanship and different levels of personal importance of the information on the perception of indication.
We also noticed that while we decided to pick the claims according to established consensus, our understanding of consensus can be imperfect. Also, further research is needed to explore the interaction effects between various content-based and context-based  credibility sources. While individual indicators have been extensively studied, a more holistic assessment considering scenarios where users encounter multiple indicators simultaneously would be valuable. 
Finally, as our primary focus was on how to present the results of AI-checking to the user, we assumed that the system predicted the information fully correctly; however, this is not always the case. Further studies should also address this fact and investigate the interfaces, which can acknowledge not just the trustworthiness of information but the certainty of the AI system itself and the way these two types of data should be jointly presented to the user. Future work should also investigate other kinds of social media posts, such as image or video posts, in particular because Generative AI makes it increasingly easy to create visual misinformation.

\section{Conclusion}

Reducing the spread of online misinformation is a critical and challenging endeavor. In this paper, we presented a preregistered online experiment (n = 537) to investigate the short way the AI-based veracity-checking system can communicate credibility assessments on social networks systematically. Specifically, our experiment compared the impacts of how indication (binary true/false or fine-grained) and different sources (content-based or context-based) on users' judgment of post veracity. We found evidence that both context-based  and content-based sources of credibility are understandable by users and can be taken into account by communicating the veracity of the information to the user. We also found that fine-grained indicators hold significant potential: they improved user information awareness, raised users' intention to use a fact-checking system, and aided in forming nuanced opinions about information through critical thinking. Our results can be used to design the presentation of AI-based credibility verification systems, including the media applying encrypted communication. 

\bibliography{Reference}
\bibliographystyle{acm}

\appendix
\section{Appendix}
\subsection{Health Post Utilized in The Study}
Below, we present the original health claim (in the title), the generated post, and the veracity of the post.

\paragraph{Post considered as False: Milk provide relief from ulcer pain}
Have you thought about using simple, natural methods to ease ulcer pain? You might be surprised to learn how effective milk can be thanks to a substance called Lactobiosealine. For a long time, people have used milk to relieve ulcer discomfort, following an old wisdom. Milk is known for being soothing, and it's thought that Lactobiosealine in milk helps calm the stomach. This substance is believed to coat the stomach lining, forming a barrier that protects against stomach acid and helps heal ulcers.While there are many modern treatments for ulcers, using something as common as milk, which supposedly contains Lactobiosealine, could be good for your health. Milk is thought to work because Lactobiosealine might help the stomach cells repair themselves, fixing the damage caused by ulcers. 

\paragraph{Post considered as False: Consuming chocolate leads to the development of acne}Recent discussions within the scientific community are bringing to the forefront a potential link between chocolate consumption and the incidence of acne, a topic that is garnering increasing interest and debate among dermatologists.
The main part of this debate is about what's in chocolate, especially its high sugar and fat content. Some people believe that these ingredients might change hormone levels in our bodies, causing acne. They say that acne often flares up when people eat more chocolate, which could support this idea.
The theory goes deeper into how chocolate's sugar and fat might affect our hormones. It suggests that these ingredients might increase hormones like insulin and androgens. These hormones can make the oil glands in our skin more active, which can lead to acne. Also, the theory looks at how certain fats in chocolate, which might cause inflammation, could make acne worse.

\paragraph{Post considered as False: Eating Eggs Can Harm Heart Health}A comprehensive meta-analysis conducted by the Institute of Organic Health and Naturopathic Science gives us a new perspective on how eggs affect heart health. The results question the old idea that eggs are always good for us, especially because of the cholesterol in egg yolks. It points out that eating eggs might increase the risk of heart problems, especially for people who already have certain types of cholesterol issues. This important discovery shows that people need to make food choices that fit their own health needs, especially if they already have health problems. This study adds a lot to the ongoing discussion about cholesterol in our diet and how it relates to heart health. It's really important to share this information so people can make better choices about eating eggs and keeping their hearts healthy.

\paragraph{Post considered as True: Screen Time Is Not a Long-Term Risk to Eyes}
Thinking about getting a big new TV but worried it might hurt your eyes? Don't worry!In our world full of screens, it's important to know how they affect our eyes. Although there's a lot of talk about this, studies show that looking at screens doesn't cause serious, lasting harm to our eyesight.The most common eye problems from using screens too much, like eye strain, dryness, and discomfort, are usually temporary and can be fixed. These issues, sometimes called Computer Vision Syndrome or Digital Eye Strain, can make your eyes feel tired, give you headaches, blur your vision, and cause neck pain. But these problems often come from things like bad lighting, screen glare, looking at the screen at a bad angle, or not taking breaks. If you take care of your eyes and use screens the right way, you can avoid these problems or even get rid of them completely.

\paragraph{Post considered as True: Dietary Supplements Not Essential for Health}
People often talk about whether taking dietary supplements is good for our overall health. There are many supplements for sale, but it's not clear if we really need them to be healthy or if we can stay healthy without them.
Most research shows that for people who don't have special health needs, supplements don't make a big difference in their overall health. Actually, relying too much on supplements can make people forget about eating healthy foods, which are the best way to get the vitamins and minerals we need. Food has a mix of vitamins, minerals, fiber, and other things that all work together to keep us healthy. Supplements don't have this mix.
Taking too many supplements can be bad for your health. Vitamins like A, D, E, and K are stored in our body and can be harmful in large amounts.

\paragraph{Post considered as True: Gut Microbiome Affects Mental Health}
Recently, scientists have become very interested in how our mental health is linked to the bacteria in our gut. They are studying the close connection between our gut and its bacteria and our mental health. The gut and brain communicate with each other in a two-way system known as the gut-brain axis. This involves our nervous system, immune system, and certain chemicals that facilitate communication between the gut and brain. Researchers are still working to understand exactly how the gut and brain affect each other. However, they have some theories. These include the production of brain chemicals like serotonin in the gut, the impact of gut bacteria on inflammation and immune function, and the role of the gut in regulating our response to stress.

    \label{fig:Example Interface}

\end{document}